\documentstyle[12pt]{article}
\newcommand{\beq}{\begin{equation}}
\newcommand{\eeq}{\end{equation}}
\newcommand{\bdis}{\begin{displaymath}}
\newcommand{\edis}{\end{displaymath}}
\newcommand{\beqa}{\begin{eqnarray}}
\newcommand{\eeqa}{\end{eqnarray}}
\newcommand{\esp}{\vspace{.1cm}}
\newcommand{\erf}{{\rm erf}}
\newcommand{\erfc}{{\rm erfc}}
\begin{document}
\begin{center}
\begin{Large}
\begin{bf}
S-Matrix Poles \\
and the \\
Second Virial Coefficient \\
\end{bf}
\end{Large}
\vspace{.7cm}
 A. Amaya-Tapia\\
 Centro de Ciencias F\'\i sicas, UNAM \\
 AP 48-3, Cuernavaca, Mor. 62251, M\'{e}xico \\
\vspace{.3cm}
S. Y. Larsen and J. Baxter \\
 Department of Physics, Temple University \\
 Philadelphia, PA 19122, USA \\
\vspace{.3cm}
Monique Lassaut \\
Groupe de Physique Th\'eorique \\
Institut de Physique Nucl\'{e}aire \\
Orsay Cedex, France \\
\vspace{.3cm}
Manuel Berrondo \\
Department of Physics, Brigham Young University \\
Provo Utah,  U.S.A. \\
\vspace{1.cm}
ABSTRACT \\
\vspace{.3cm}
\end{center}
\baselineskip=14pt
For cutoff potentials, a condition which is not a limitation for
the calculation of physical systems,
the S-matrix is meromorphic. We can express it in terms of
its poles, and then calculate the quantum mechanical second
virial coefficient of a neutral gas.

Here, we take another look at this approach, and discuss the feasibility,
attraction and problems of the method. Among concerns are the rate of
convergence of the `pole' expansion and the physical significance of the
`higher' poles.

\newpage
\topmargin=0in
\pagestyle{plain}
\pagenumbering{arabic}
\setcounter{page}2
\section{ Introduction}
\vspace{.3cm}

\vspace{.3cm}

  This work has had a long gestation. Our earliest notes date from
July 1976! At that time, two of us (M. Berrondo and S.Y. Larsen)
 obtained what we thought was an attractive
formula for the q.m. second virial coefficient in terms of the location
of its poles, as an alternative to its well known\cite{uhl} and customary
formulation in terms of phase shifts and bound state energies.
Soon after doing this work, and obtaining some results, we were made
aware that H. Nussenzveig had already published\cite{nuss1} an attractive
article on the same subject and we stopped our efforts.

Recently, however, there has been interest in some other quarters\cite{sof},
related to the pole expansion of the S-matrix, and we thought that useful
details of such expansions might be properly be brought to light within
the context of the statistical mechanics of the virial.
Together with other colleagues, we look at the rate of convergence for
phase shifts and  for the virial. We look at asymptotic expansions for the
poles.
We try to see whether we can use tricks to either accelerate the convergence
or sum background poles.
We look at different model potentials. In the case of hard spheres, we obtain
an expansion for the virial valid at low temperature.

\section{The second virial in terms of the poles}

For potentials decreasing faster than exponentials, like gaussian or cutoff
potentials the Jost function is an entire function and, accordingly, 
the ${\cal S}$ matrix is meromorphic. We can express it as a exponential  together with
a product of poles :

\beq
{\cal S}_{\ell} (k) = e^{2 i \delta_{\ell}} =
(-)^{\ell} \ e^{-2 i k a} \ \prod_n
\frac{k_{n,\ell}+k}{k_{n,\ell}-k} \ ,
\eeq

\noindent where $a$ denotes the range of the potential and the
$k_{n,\ell} $'s are the  poles
labelled in order  of increasing modulus.   It should be noted that this
expression, apart from the factor $(-)^l$, which we will examine later,
 is equivalent to the one  derived some years ago by Nussenzveig\cite{Nuss}
\beq
{\cal S}_{\ell} (k) =  \ e^{-2 i k a} \ \prod_n
\frac{k_{n,\ell}^*-k}{k_{n,\ell}-k} \ .
\eeq
We recall that the original\cite{beth} quantum mechanical formulation of the
Boltzmann part of second virial coefficient, in term of phase shifts and
bound state energies,  reads
\beq
\left( B_2 \right)_{Boltz} = -2^{1/2}  \lambda_T^3 {\cal N}    \left[
 \sum_{B,\ell} (2 \ell +1)
e^{- 2  \beta  k^2_{B,\ell}} + \frac{1}{\pi} \ \int_0^{\infty}
 dk \ e^{-2  \beta k^2} \ \sum_{\ell}
 (2 \ell +1 ) \frac{d}{dk} \delta_{\ell} (k)
 \right] \ ,
\eeq
 where $\beta =1/(\kappa  T)$ in terms of the Boltzmann 's constant $\kappa$
 and the
temperature $T$, $ \lambda_T$ denotes
$h/\sqrt{ 2 \pi m \kappa T} $ and ${\cal N}$ is the number of particles in
the volume $V$. (Note that $2 \beta = \lambda_T^2 /(2 \pi)$).
The sum for the  bound states runs over $\ell$ and the number of bound states
for each $\ell$.

\newpage

In later calculations and formulations\cite{BLK}, a partial integration
 has often
been performed to yield the virial in terms of the phase shifts,
themselves, and the bound state energies.
Also for convenience, we have divided the virial into two parts, a Boltzmann
 plus an exchange part. We focus on the Boltzmann part,
 but can at any moment obtain the exchange part by
 minor internal changes of sign and the inclusion of a perfect gas term.

Since the ${\cal S}$ matrix, here,  has a compact expression in
terms of its poles the previous equation also has such an expression.
Indeed
the derivative $d \delta_{\ell}/dk $ is nothing else than
\beq
\frac{d}{dk} \delta_{\ell} =-a +
\frac{1}{2 i} \sum_n \left[ \frac{1}{k_{n,\ell}+k}
+\frac{1}{k_{n,\ell}-k} \right] \ .
\eeq

Introducing (4) in (3) we look at the contribution from the bracket
\beq
\sum_{B,\ell} (2 \ell +1)  e^{- 2  \beta  k^2_{B,\ell}} +
\frac{1}{\pi} \ \sum_{l} (2 \ell +1) \left(- \frac{a}{2} \sqrt{\frac{\pi}{2
\beta}}
 - \sum_n \frac{1}{i} \ \int_0^{\infty}
 dk \ e^{-2  \beta k^2} \ \frac{k_{n,\ell}}{k^2-k^2_{n,\ell}}
  \right) \ .
\eeq
The integral involving the  pole expansion
 can be written in terms of the error function
\beq
 \erf(z)=\frac{2}{\sqrt{\pi}} \int_0^z e^{-v^2} dv \ .
\eeq
In Appendix A we show that
\beq
- \frac{ k_{n,\ell}}{i} \int_0^{\infty}
 dk \ e^{-2  \beta k^2} \ \frac{1}{k^2-k_{n,\ell}^2}
 = \frac{\pi}{2} e^{-2 \beta k_{n,\ell}^2}
 \left(\erf(-i k_{n,\ell} \sqrt{2
 \beta} ) \mp 1 \right) \ ,
\eeq
with the sign - when the $\Im(k_{n,\ell}) $ is positive and the sign +
when the $\Im(k_{n,\ell}) $ is negative.

\noindent The bracket then reads
\beq
\sum_{B,\ell} (2 \ell + 1) e^{- 2  \beta  k^2_{B,\ell}} +
\ \ \sum_{l} (2 \ell +1) \left(- \frac{a}{\sqrt{2} \sqrt{ 4 \pi \beta}}
+ \frac{1}{2} \sum_n   e^{-2 \beta k_{n,\ell}^2}
 \left[ \erf(-i k_{n,\ell} \sqrt{2  \beta})
 \mp 1 \right] \right) \ .
\eeq
We now put together the terms corresponding to bound states ($\sum_{B,\ell}$)
and the terms corresponding to bound states, in the expression involving 
the poles ($\sum_{n,\ell}$). 
The poles involved in these terms are situated in the upper
half plane $\Im(k)>0$.
According to our previous discussion,  the exponential
$\exp(- 2 \beta (k_{n,\ell})^2) $ is then weighted by the
factor (-1/2). Combined with the exponential terms of the bound states,
this yields the factor +1/2 as for the other poles. The bracket then reads
\beq
\sum_{l} (2 \ell +1) \left( - \frac{a}{ \sqrt{2}
 \sqrt{ 4 \pi \beta}}  \
 + \frac{1}{2}  \sum_n
  e^{-2 \beta k_{n,\ell}^2}
(\erf(-i k_{n,\ell} \sqrt{2  \beta})+1) \right) \ .
\eeq
Remembering, now, that $\erf(-z)=-\erf(z)$ and $1-\erf(z)=\erfc(z)$, 
and using
$ 2 \beta = \lambda_T^2/(2 \pi)$ we  obtain our final expression ( in terms of
$\lambda_T$):
\beq
\left( B_2 \right)_{Boltz} = - 2^{1/2} \lambda_T^3 {\cal N} \
 \sum_{l} (2 \ell +1)
\left[ -\frac{a}{ \sqrt{ 2 } \lambda_T}
+ \frac{1}{2}  \sum_n \exp(- \frac{\lambda_T^2}{2 \pi}
k_{n,\ell}^2)
\erfc(i \frac{ \lambda_T}{\sqrt{2 \pi}}  k_{n,\ell}  )   \ \right] \ .
\eeq

\section{The phase shifts in terms of poles}

Clearly our procedure becomes more attractive if few poles are required
 to reproduce the
second virial to good  accuracy. In this section, we examine how
the expansion, Eq.(4),  reproduces the phase-shifts.

For a pure hard sphere potential, the answer is very pleasing: the number of
poles for the $\ell^{\rm th}$partial wave equals $\ell$.

For $\ell = 0$, the phase
shift of a hard sphere of radius $\sigma$ is just $- k\sigma$ and the poles
do not contribute. For the higher $\ell$'s, we find that our formula for the
S-matrix works perfectly, but that the factor $(-1)^\ell$ is not needed.
We note that asymptotically  the phases tend to $- k\sigma - \ell\pi/2$.

Unfortunately, for the more common cutoff potentials, the number of poles
appearing in the expansion is infinite and, practically, to obtain the phase
shift to, say 5 digits accuracy, the number of poles required is impressively
large. We note, though, that except for poles found on the imaginary axis,
they occur in pairs, in the third and fourth quadrant, and thus for these
pairs, it is sufficient to determine the poles in the last quadrant.
For a potential made up of a (repulsive) hard core plus an
attached attractive square well, we find the results presented in table 1.

\esp
\begin{center}
\begin{tabular}{|l|r|c|}
\hline
N & $ \delta_0$ \ \  & ${\cal S}_0$ \ \ \  \\
\hline
22     & .067181  & (.9909869, .1339585) \\
72     & .073081  & (.9893373, .1456426) \\
122    & .074379  & (.9889561, .1482090) \\
522    & .076023  & (.9884633, .1514605) \\
1022   & .076312  & (.9883756, .1520316) \\
2022   & .076473  & (.9883265, .1523510) \\
10022  & .076620  & (.9882818, .1526403) \\
50022  & .076654  & (.9882713, .1527083) \\
150022 & .076661  & (.9882693, .1527210) \\
\hline
 Exact & .076664 &  (.9882682, .1527283) \\
\hline
\end{tabular}
\end{center}
\begin{center}
{\it Table 1.  Poles for 1 antibound state + $N$ pairs}
\end{center}
\esp

\noindent This is for $\ell =0$, $k \sigma=q =0.1$,
 the de Boer parameter $\Lambda^*=(h^2/m V_0 \sigma^2)^{1/2}$
equal to $10$, the hard core radius $\sigma$ and finally $a$, the
outer limit  of the attractive potential, equal
to $2.85 \sigma$.

The expansion converges but so slowly that  we are faced with the
necessity of increasing the rate of convergence of the series.
We note that at the origin $(k=0)$, the phase shift behaves like
$k^{2 \ell + 1} $. Thus, for a given $\ell$, derivatives up to $2 \ell$
are equal to zero. We obtain:

\beqa
-a + \frac{1}{i} \sum_n \frac{1}{k_{n,\ell}} & = 0 & \rm for \ \ \ell \geq 1
\nonumber \\
 \sum_n \frac{1}{k_{n,\ell}^{2 j + 1 }} & = 0 &
 1 \leq j \leq \ell-1
\ \rm for  \ \ \ell \geq 2 \ .
\eeqa
We can now try to accelerate the convergence of (4) by subtracting the
terms above, multiplied by appropriate powers of $k$. We obtain, for $\ell
\geq 1$,
\beq
\frac{d}{dk} \delta_{\ell} = \frac{1}{2 i} \sum_n \left[
\frac{1}{k_{n,\ell}+k}
+\frac{1}{k_{n,\ell}-k} -\frac{2}{k_{n,\ell}}  \sum_{j=0}^{\ell-1} \left(
\frac{k}{k_{n,\ell}}\right)^{2 j}  \right]  \ ,
\eeq
which can also be written as 
\beq
\frac{d}{dk} \delta_{\ell} = \frac{1}{ i} \sum_n
\left( \frac{k}{k_{n,\ell}} \right)^{2 \ell } 
\ \frac{k_{n,\ell}}{k_{n,\ell}^2 -k^2} \ .
\eeq
We see that, for high orders, the terms behave as $1/(k_{n,\ell})^{ 2\ell+3}$.

Integrating the above equations with respect to $k$ we find
\beq
\delta_{\ell}(k) = \delta_{\ell}(0) + \frac{1}{2 i}  \sum_n 
\left[ \ln \left(\frac{k_{n,\ell}+k}{k_{n,\ell}-k}\right) 
- \sum_{j=0}^{\ell-1} \frac{2}{2 j + 1}
 \left(\frac{k}{k_{n,\ell}}\right)^{2 j + 1} \right] \qquad\quad  \ell \geq 1
 \ .
\eeq
For  $\ell=1,q=0.1$ and 23 pairs of poles, we obtain
$1.478248 \ 10^{-3}$ compared to the exact result of $1.47826613 \ 10^{-3}$
and, similarly, for $\ell=2$,
$4.04298921 \ 10^{-6} $ instead of $4.042988889 \ 10^{-6} $ .
The method, however, deteriorates as the energy increases.

To remedy the slow convergence for $\ell=0$, we
introduce the derivative of the phase
shift (non zero) at $k=0$ , i.e. 
\beq
\frac{d}{dk} \delta_{0}(k) = \frac{d}{dk} \delta_0(k) \vert_{k=0} +
\frac{1}{2 i} \sum_n \left[ \frac{1}{k_{n,0}+k}
+\frac{1}{k_{n,0}-k} - \frac{2}{k_{n,0}} \right] .
\eeq
We then easily calculate the new term using trigonometric functions.

To prevent the deterioration as $k$ increases, we limit the number of 
subtractions to 2. 
The results are then presented in table 2. 
Clearly, the number of poles required,
 $\ell$ being fixed, rises as the energy increases. On the other hand,
 at fixed energy, the number of poles needed is least when $\ell$ is largest.

\esp
\begin{center}
\begin{tabular}{|l|r|r|r|r|r|r|r|}
\hline
$k \sigma$ \ $\ell$ & 0 & 1 & 2 & 3 &4 & 5 &6 \\
\hline
 1 &    69  &  70  &  19   &  20  &  21  &  22 &    23 \\
 2 &    141 &  142  &  37  &  38  &  39  &  40 &   41 \\
 3 &   217 &  218  &  55 &   56  &  57 &   58  &  59 \\
 4 &   291 &  292  &  73 &   74  &  75 &   76 &   77 \\
 5 &   367 &  368  &  91 &   92  &  93 &   94  &  95 \\
 6 &    443 &  444  & 111 &   112 &  113 &  114 &  115 \\
 7 &    519 &  520  & 129 &  130 &  131 &  132 &  133 \\
 8 &   597 &  598  & 147  & 148  & 149  & 150  & 151 \\
 9 &    673 &  674 &  165 &  166 &  167 &  168  & 169 \\
 10 &   751 &  752  & 185 &  186 &  187 &  188 &  189 \\
\hline
\end{tabular}
\end{center}
\begin{center}
{\it Table 2. Poles required for 5 digit accuracy}
\end{center}
\esp

\section{Asymptotic formulas}
The simplest way to remedy a lack (or a limited number) of poles is
to determine an asymptotic formula. Nussenzveig, in his book\cite{Nuss},
derived  such an asymptotic expression, which we extended to `handle ' 
hard cores,
but we need to increase its accuracy.

We begin by looking for an asymptotic formula for the $s$-wave, and then 
generalize to higher angular momenta.
The zeros of the Jost function $F(k)$ are the poles of the S-matrix
$F(-k)/F(k)$. They are given (see Appendix B) by solving the equation
\beqa
e^{2 i k (a-\sigma) } & = & \frac{4 k^2}{V(a)} 
 \left[ 1 + \frac{i}{2 k}  \left(2 M-\frac{V'(a)}{V(a)}\right) \right. 
 \nonumber \\
& +& \left. \frac{1}{4 k^2} \left(\frac{V''(a)}{V(a)}-\frac{V'(a)^2}{V(a)^2}
- 2 V(a) + 
2 M \frac{V'(a)}{V(a)} - 2 M^2  \right) \right]  \ ,
\eeqa
where $M$ denotes $M=\int_{\sigma}^a V(r') \ dr'$ (for a potential finite
at the origin put $\sigma=0$) and $V(a)$ is assumed different from
zero.

\noindent We thus obtain  some  corrections to Nussenzveig's 
leading term, which were calculated from the equation:
\beq
e^{2 i k (a-\sigma) } = \frac{4 k^2}{V(a)} \ .
\eeq
To get the correct value we first solve  equation (16) by setting
\beq
k \ (a-\sigma) = n \pi -\epsilon \ \pi/2 - i \ \Delta \ ,
\eeq
where $\epsilon=0,1$ according to whether  $V(a)$ is positive or
negative, and, then iterating
\beq
e^{ \ \Delta} = \frac {2(x_0 - i \Delta)}{A}  \ ,
\eeq
with $x_0=n \pi -\epsilon \ \pi/2 $ and $A^2 = \vert V(a) \vert (a-\sigma)^2 $.
The first correction is
\beq
\Delta_0=    \ln \frac{2 x_0}{A} \ ,
\eeq
which corresponds to that of Nussenzveig. The third iteration, practically,
provides us with the exact value so that
\beq
\Delta \simeq \Delta_2 =  \ln \left[ \left(
2 x_0 -2 i \ln((2 x_0-2 i \ln(2 x_0 /A))/A) \right)/A \right] \ .
\eeq
To get higher order corrections we have to include additional  terms in
(17), involving higher powers of $1/k$, and solve by iteration
\beq
e^{ \ \Delta} = 2 \  \frac {x_0 - i \Delta}{A} 
\left[ 1 + \frac{\alpha_1}{x_0 - i
\Delta} + \frac{\alpha_2}{(x_0 - i \Delta)^2} +\cdots \cdots \right] \ ,
\eeq
with
\beqa
 \alpha_1 & = & \frac{i (a-\sigma)}{4}  
 \left(2 M-\frac{V'(a)}{V(a)}\right) \nonumber \\
\alpha_2 & = &  \frac{(a-\sigma)^2}{8} \left( \frac{V''(a)}{V(a)}
 -\frac{3}{4} \ \frac{V'(a)^2}{V(a)^2} -2 V(a) + M \frac{V'(a)}{V(a)}
 -M^2 \right)  \ .
\eeqa
To test the method for the potential that we used before ( hard core +
 square well), we define $A = 2 \pi (a-\sigma)/\Lambda^*$.
An  expansion in $1/k$ up to the $4^{th}$ order yields the results in 
table 3.
We see that we are  able to reproduce the exact results, for the higher poles,
to  5  digits.

\esp
\begin{center}
\begin{tabular}{|r|r|c|}
\hline
       &   n     &   $k_{n,0}$ \\
\hline
 appr. &           20 &      (33.0670568,-2.5190331) \\
 exact          &   20  &(33.0670586,-2.5190330) \\
\hline
 appr. &           21 &   (34.7670884,-2.5459707) \\
 exact &           21 &   (34.7670900,-2.5459707)  \\
\hline
 appr. &           22 &   (36.4669649,-2.5716312) \\
 exact     &      22  &   (36.4669647,-2.5716312) \\
\hline
 appr. &           23 &   (38.1667055,-2.5961305) \\
 exact &          23 &    (38.1667061,-2.5961304) \\
\hline
\end{tabular}
\end{center}

\begin{center}
 {\it Table 3. \ \ $\Lambda^*=10$, \  $a/\sigma = 2.85$}
\end{center}
\esp

\noindent For waves with higher values of $\ell$,
 we show, in Appendix C, that we can generalize the approach of Appendix B, for
 example by setting  the free Jost solution equal to 
\beq
 w_{\ell}(k r) = i^{\ell + 1} \sqrt{ \frac{\pi}{2}  k r} \ H^{(1)}_{\ell +
 1/2}(k r) \ ,
\eeq
and proceeding in a manner similar to that used for $\ell=0$.

 In table 4, we examine the quality of our approximations for values 
of $\ell$ not equal to zero. Our results for  
 $\ell=1,2$ are  of comparable quality to  those that
we found   for the $l=0$.
The key parameter is the ratio $\ell/k(a-\sigma)$. So long as it is small the
asymptotic results will be good.
We see that for large  $\ell$'s, we have to proceed to a larger value of
$n$, before the behaviour of the poles becomes asymptotic.

\esp
\begin{center}
\begin{tabular}{|r|r|c|}
\hline
  $\ell$  & &   $k_{23,\ell}$ \\
\hline
1 & appr. &       (38.1758622,-2.5955925) \\
 &  exact    &    (38.1758614,-2.5955925) \\
\hline
2 &  appr. &       (36.4956875,-2.5698867) \\
  & exact  &      (36.4957008,-2.5698843) \\
\hline
3 &  appr. &       (36.5243519,-2.56815261) \\
  & exact  &      (36.5244255,-2.56813956) \\
\hline
4 &  appr. &       (34.8672057,-2.53969925) \\
 & exact      &  (34.8674660,-2.53965187) \\
\hline
5 &  appr. &       (34.9169320,-2.5366040) \\
 & exact &       (34.9175873,-2.5365038) \\
\hline
6 &  appr. &       (33.2864742,-2.5046694) \\
 & exact &       (33.2883530,-2.5045969) \\
\hline
\end{tabular}
\end{center}

\begin{center}
{\it Table 4. \ \ $\Lambda^*=10$, \  $a/\sigma = 2.85$}
\end{center}
\esp

When the potential has no hard core, we show, in Appendix C,  
 that the asymptotic expression of the poles is obtained (in lowest order)
from the solution of
\beq
e^{2 i k a} = (-)^{\ell} \frac{V(a)}{4 k^2} \ .
\eeq
We then recover the $(-)^{\ell}$ dependence,
 mentioned earlier by Nussenzveig\cite{Nuss}.
This dependence disappears when the potential incorporates a hard core.

\section{Virial}
Noting, in the previous section, the necessity of accelerating the convergence
of the pole expansion, we present here the formalism for doing
this for the virial. Afterwards we will discuss the application to 
various potentials.

Given the slowness of the basic expansion in terms of poles, we modify the
basic virial equations, proposing two different versions. In the first one 
we simply write 
\beq
\left( \frac{d}{dk} \delta_{\ell}(k) \right)_{k=0} = -a + \frac{1}{i} \ \sum_n
 \frac{1}{k_{n,\ell}}  \ ,
\eeq
and add and subtract this from the derivative expression. This yields:
\beqa
\left( B_2 \right)_{Boltz} & = & - 2^{1/2} \lambda_T^3 {\cal N} \
\left[ \frac{1}{2^{1/2} \lambda_T} 
\ \left( \frac{d}{dk} \delta_{0}(k) \right)_{k=0} \right. \nonumber\\  
 & + & \left. \frac{1}{2} \sum_{l,n} (2 \ell +1) 
\left( \exp(- \frac{\lambda_T^2}{2 \pi}
k_{n,\ell}^2) \ \ 
\erfc(i \frac{ \lambda_T}{\sqrt{2 \pi}}  k_{n,\ell}  )
+ \frac{2^{1/2} i}{k_{n,\ell} \lambda_T} \right) \ \right] \  .
\eeqa
We have used the fact that 
\beq
\left( \frac{d}{dk} \delta_{\ell}(k) \right)_{k=0} = 0 \ \ \rm for \ \ \ell >0
 \ .
\eeq
We can push this further, using Eq.(11) of our paper. We then find that 
\beqa
\left( B_2 \right)_{Boltz} &  = & - 2^{1/2} \lambda_T^3 {\cal N} \
\left[ \frac{1}{2^{1/2} \lambda_T} 
\ \left( \frac{d}{dk} \delta_{0}(k) \right)_{k=0} \right. \nonumber\\  
& + & \left. \frac{1}{2} \sum_{l,n}  (2 \ell +1) 
 \exp(- \frac{\lambda_T^2}{2 \pi}
k_{n,\ell}^2) \ 
\erfc(i \frac{ \lambda_T}{\sqrt{2 \pi}}  k_{n,\ell}  ) \right. \\ 
& + & \left. \frac{ i}{2^{1/2} \lambda_T} \left(
\sum_{\ell,n} (2 \ell+1) \frac{1}{k_{n,\ell}} + \sum_{\ell = 2}^\infty
(2 \ell + 1)  \
\sum_{n}
\sum_{j=1}^{\ell-1} \  (2 j - 1) !! \ (\frac{\pi}{\lambda_T^2})^j \
\frac{1}{k_{n,\ell}^{2 j +1}} \right)  
  \right] \ . \nonumber 
\eeqa
\esp
Finally, what we calculate is this virial divided by that 
obtained classically for a pure hard core of radius $\sigma$, i.e.

\beq
\left( B_2^* \right)_{Boltz} = \frac{3}{2 \pi {\cal N} \sigma^3}
\ \left( B_2 \right)_{Boltz}.
\eeq

\subsection{Hard spheres}

We illustrate in table 5, how with a limited number of poles
(about 20) we reproduce the $2^{nd}$ virial coefficient for hard spheres,
over a range of relatively low temperatures, for which these formalisms
are designed.

\esp
\begin{center}
\begin{tabular}{|c|c|c|c|c|}
\hline
 & \multicolumn{2}{c|}{$\lambda_T/\sigma=1$} &
 \multicolumn{2}{c|}{$\lambda_T/\sigma=30$}   \\
\hline
$\ell$ & BLK\cite{BLK} & Poles & BLK\cite{BLK} & Poles   \\
\hline
0 & .477464829276 & .477464829276 & .429718346348E+03  & .429718346348E+03  \\
1 & .752348708365 & .752348708365  & .445367908873E+01  & .445367908873E+01 \\
2 & .584467004923 & .584467004923  & .867588096488E-02  & .867588096488E-02 \\
3 & .335935832978 & .335935832978 & .848769265246E-05  & .848769265246E-05 \\
4 & .155466787903 & .155466787903 & .544399040752E-08  & .544399040752E-08 \\
5 & .059852912973 & .059852912973 & .258124235604E-11
& .258124235604E-11 \\
\hline
\end{tabular}
\end{center}
\begin{center}
{\it Table 5. The hard sphere virial, from Eq.(29)}
\end{center}
\esp

In Appendix D, we outline how we recover terms in a low temperature
expansion ($\sigma/\lambda_T$ small), 
a number of which are found in BLK.

\subsection{Hard sphere + square well}

For $\Lambda^{*} = 4$, the $2$ particles have a bound state for $\ell = 0$ and
one for $\ell = 1$. These bound states then dominate the very low temperature 
behaviour of these partial virials (i.e. for these angular momenta), and 
therefore for these virials. This is due to the large exponential terms
that arise for large $\lambda_T/\sigma$. This already becomes evident for 
$\lambda_T/\sigma = 2$, and becomes more important for $\lambda_T/\sigma$ =
$5$, $10$, $20$, $30$ $\cdots$ 

Still, for $\lambda_T/\sigma$ = $2$, it is useful to see the virial
calculated and displayed as
function of the number of poles taken into account, for different values of
 $\ell$. In table 6 we subtract the first derivative as in Eq. (27). 

Here it is well to comment on our reference BLK. The authors BLK published
results on hard spheres\cite{BLK}, many years ago, but not partial results
for distinct values of $\ell$, nor results for a hard sphere + a square well.
Here,  by BLK, we mean that
one of the previous authors (SYL) is using some of the old programs and the
old methods
to obtain results which can be used to benchmark the use of the poles.
These results, however,  have not been obtained with the same need and desire
for accuracy that the old work required. The programs have now been used with
machines
with smaller word lengths, the calculations done with larger meshes, etc.
The values of the bound states have also been obtained from our work with
the poles.

\esp
\begin{center}
\begin{tabular}{|l|c|c|c|c|}
\hline
N $\ell$ & 0   &                1  &           2    &         3    \\
\hline
B+AB  & -4.96245020 & -13.49991189  &   9.568499294   &         \\
\hline
   1  & -5.38351004 & -14.59384163 & -13.71610648  & 7.071340875 \\
   2 &  -5.42559329 & -14.70879231 & -15.01832365 &  -8.916442069 \\
   5 &  -5.44089652 & -14.75294247 & -15.23836884 & -10.22935288 \\
  10 &  -5.44304631 & -14.75932474 & -15.25610946 & -10.27349937 \\
  20 &  -5.44347274 & -14.76060000 & -15.25890823 & -10.27875744 \\
  50 &  -5.44355258 & -14.76083933 & -15.25936689 & -10.27949972 \\
 100 &  -5.44355890 & -14.76085827 & -15.25940041 & -10.27954958 \\
 200 &  -5.44355982 & -14.76086106 & -15.25940520 & -10.27955649 \\
 300 &  -5.44355993 & -14.76086137 & -15.25940573 & -10.27955725 \\
 400 &  -5.44355996 & -14.76086145 & -15.25940587 & -10.27955744 \\
\hline
BLK &
\multicolumn{1}{l|}{-5.44355995} &
\multicolumn{1}{l|}{-14.7608616} &
\multicolumn{1}{l|}{-15.2594062} &
\multicolumn{1}{l|}{-10.2795575} \\
\hline
\end{tabular}
\end{center}
\begin{center}
 Table 6. \ \ $\Lambda^*=4$, \ $a/\sigma = 2.85$,\  $\lambda_T/\sigma=2.$ \\

{\it The virial as a function of the number N of pairs,
for different angular momenta.
The contribution of bound (B) and antibound
(AB) states is given separately at the beginning of the table.}

\end{center}
\esp

Here it is well to comment on our reference BLK. The authors BLK published
results on hard spheres\cite{BLK}, many years ago, but not partial results
for distinct values of $\ell$, nor results for a hard sphere + a square well.
Here,  by BLK, we mean that 
one of the previous authors (SYL) is using some of the old programs and the
old methods
to obtain results which can be used to benchmark the use of the poles.
These results, however,  have not been obtained with the same need and desire
for accuracy that the old work required. The programs have now been used with
machines
with smaller word lengths, the calculations done with larger meshes, etc.
The values of the bound states have also been obtained from our work with
the poles.

This said, we present two tables, exclusively with poles,
for $\Lambda^{*} = 10$ and $a/\sigma = 2.85$, for which
there are no bound states. In table 7 we show, for $\ell = 0$,
the convergence of the
virial as a function of the number of pairs, for different values of 
$\lambda_T/\sigma$.

\esp
\begin{center}
\begin{tabular}{|l|c|c|c|c|}
\hline
  N $\lambda_T/\sigma$ & 1 & 2 & 5 & 10 \\
\hline
AB    &  0.598187603 & 1.554449992 & 0.649915614 & -20.719592323 \\
\hline
   1 &  0.503078797 & 1.357448734 & 0.454501918 & -20.909727874 \\
   2 &  0.457717502 & 1.318856199 & 0.420668117 & -20.942948517 \\
   5 &  0.434770286 & 1.301162611 & 0.404123797 & -20.959343787 \\
  10 &  0.431866213 & 1.298475743 & 0.401489560 & -20.961970746 \\
  20 &  0.431323513 & 1.297945047 & 0.400962083 & -20.962497768 \\
  50 &  0.431225758 & 1.297847856 & 0.400865048 & -20.962594781 \\
 100 &  0.431218271 & 1.297840377 & 0.400857571 & -20.962602258 \\
 200 &  0.431217192 & 1.297839297 & 0.400856492 & -20.962603337 \\
 300 &  0.431217071 & 1.297839177 & 0.400856371 & -20.962603457 \\
 400 &  0.431217040 & 1.297839146 & 0.400856349 & -20.962603488  \\
\hline
\end{tabular}
\end{center}
\begin{center}
{\it Table 7. \ \  $\Lambda^*=10$, \  $a/\sigma=2.85$, \  $\ell = 0$}
\end{center}
\esp

In the table 8, we sum over the angular momenta, up to $\ell = 7$,
 noting that the largest contribution for $\ell =7$ is of the order of
 $10^{-7}$, and show the
convergence of the virial as a function of the number of pairs per $\ell$, for 
different temperatures.

\esp
\begin{center}
\begin{tabular}{|l|r|r|r|r|}
\hline
  N $\lambda_T/\sigma$ &
\multicolumn{1}{c|}{1} &
\multicolumn{1}{c|}{2} &
\multicolumn{1}{c|}{5} &
\multicolumn{1}{c|}{10} \\
\hline
AB &  4.19599724 &       9.11116433 &       13.85909885
&  -5.12661171 \\
\hline
 1  &  10.80533807  &  15.47507182 &   8.97472495 & -18.26521543  \\
 2  &  11.51743916  &  10.90069896 &  -0.65662473 & -28.69768995   \\
 5  &   2.21953160  &   0.10665942 & -10.39425326 & -38.23655996 \\
 10 &   1.33830379  &  -0.58575152 & -11.05109579 & -38.88877474 \\
 20 &   1.25987331  &  -0.66118370 & -11.12575192 & -38.96332180 \\
 50 &   1.25055050  &  -0.67043615 & -11.13498509 & -38.97255257 \\
 100&   1.24998520  &  -0.67100089 & -11.13554949 & -38.97311816 \\
 200&   1.24991013  &  -0.67107586 & -11.13562419 & -38.97319073 \\
 300&   1.24990207  &  -0.67108401 & -11.13563272 & -38.97319829 \\
 400&   1.24990001  &  -0.67108599 & -11.13563474 & -38.97320208 \\
\hline
\end{tabular}
\end{center}
\begin{center}
{\it Table 8. The complete Boltzmann virial}
\end{center}
\esp

\subsection{Further comments on the poles}

As we saw, if the discontinuity in the potential at the cut-off $a$ is not
zero, and also not infinite, then our asymptotic expansions,
for the location of the poles which appear in the expression of the S-matrix,
show that the number of these poles is infinite.

Alternatively, we can use an elegant argument, of
Newton\cite{Newt} and Nussenzveig\cite{Nuss}, which argues that the function
$G(k)=F(k) F(-k)$, involving the Jost function $F(k)$, is an
entire function of the variable $k^2$, of order 1/2,
and therefore has an infinite number of zeroes, which then again leads to
the conclusion that the S-matrix has an infinite number of poles.
The basic element in both approaches is an analysis of the behaviour of
the function $G(k)$ for large values of $|k|$. 

As we can see directly, from our asymptotic expressions, the zeroes of
$F(k)$, for these large values of $|k|$, depend on the value of the potential
at $a$, rather than on the values for $r$ from $0$ to $a$.
Thus, as already noted by Nussenzveig, the resulting poles do not have much
physical significance. He shows that a Yukawa-type potential yields a branch
cut, which if the potential is cut off, is replaced by an infinite set
of poles.

One further point. We can readily understand that some of the poles 
represent bound states and resonances. In general, it is subtle to understand
the physical significance of the poles. Nussenzveig dedicates a chapter
in his book (Causality and Dispersion Relations) to this purpose.
We commend it to our readers.

\section{Conclusion}

We think that the development of a formalism for the 
second virial coefficient, in terms of the poles of the S-matrix, is an
attractive one.
It is a formalism equally as
powerful as the more conventional one, based on phase shifts and bound states,
but, for example, treats the phase shifts and bound state contributions in 
a unified way.

For hard spheres, the number of poles for each $\ell$ is finite and we see
that, with a handful of these poles, we can reproduce results obtained
by previous methods.

For other more realistic potentials, and our hard
sphere + square well, the number of poles is infinite and 
the phase shifts and the virial converge slowly in terms of the poles.
We have, however, been able to devise tricks to accelerate this convergence. 
We have also perfected and extended the use of asymptotic expressions for 
the location of the poles.
The result is that we can still obtain (and have obtained) results with a
modest (non forbidding) number of poles. 

These days, with the abundance of numerical power available,
and the new methods that have been developed to locate poles\cite{sof}
for realistic potentials, we are freer to choose the methods that we might
use for virial calculations.

\section*{Acknowledgments}
Two of us (A. A-T and S.Y.L.) gratefully acknowledge the warm hospitality 
of the IPN-Orsay (IN2P3-CNRS) where an important part of this work was 
carried out. We thank K. Chadan for judicious comments.
Amaya is grateful for the support from the CONACYT under its grant 32175-E,
and Larsen thanks the Centro de Ciencias F\'\i sicas, 
UNAM, which has, many times in the past, extended its welcome.

\newpage
\section*{Appendix A}
The integrals in (5) are of the type
$$ 
\ \int_0^{\infty}
 dk \ e^{-2  \beta k^2} \ \frac{k_n}{k^2-k_n} \ ,
\eqno (A1) $$ 
where we drop, for convenience, the subscript $\ell$ in the expression
of the $k_{n,\ell}$'s.

Now,let
$$
I(\lambda) = e^{-\lambda k_n^2}
\ \int_0^{\infty}
 dk \ e^{-\lambda k^2} \ \frac{1}{k^2-k_n^2} \ .
\eqno (A2) $$
The function $I(\lambda)$ satisfies
$$
\frac{d}{d \lambda} I(\lambda) = -e^{\lambda k_n^2} \frac{\sqrt{\pi}}{2
\sqrt{\lambda}} \ .
\eqno (A3) $$
We have therefore
$$
I(\lambda) = I(0) - \frac{\sqrt{\pi}}{2}
\int_0^{\lambda} \ d \lambda' e^{\lambda' k_n^2} \frac{1}{
\sqrt{\lambda'}} \ .
\eqno (A4) $$
Introducing
$$
\erf(z)=\frac{2}{\sqrt{\pi}} \int_0^z e^{-v^2} dv \ ,
\eqno (A5) $$
we have
$$
I(\lambda) = \int_0^{\infty} dk \frac{1}{k^2-k_n^2}  - \frac{\pi}{2 k_n}
\erf(- i \sqrt{2 \beta k_n}) \ .
\eqno (A6) $$
By using the residues, note that
\beqa
\int_0^{\infty} dk \frac{1}{k^2-k_n^2} & = & + \frac{\pi i}{2 k_n} \ \Im(k_n)
>0 \nonumber \\ 
 & & - \frac{\pi i}{2 k_n} \ \Im(k_n) < 0 \ , \nonumber 
\eeqa
and thus
$$
- \frac{k_{n,\ell}}{i} \int_0^{\infty}
 dk \ e^{-2  \beta k^2} \ \frac{k_{n,\ell}}{k^2-k_{n,\ell}^2}
 = \frac{\pi}{2} e^{-2 \beta k_{n,\ell}^2} 
 \left(\erf(-i k_{n,\ell} \sqrt{2
 \beta} ) \mp 1 \right) \ ,
\eqno (A7) $$
with the sign - when $\Im(k_{n,\ell}) $ is positive and the sign + 
when $\Im(k_{n,\ell}) $ is negative.

\newpage
\section*{Appendix B}
Here, we derive an asymptotic expression  for the location of the poles
of the ${\cal S}$ matrix,or, equivalently of the zeros of the Jost function
(the zeros of the Jost function $F(k)$ are the poles of the S-matrix
$F(-k)/F(k)$). We solve a  Volterra  equation, which
generates the Jost solution $f_{\ell}(k, r)$
 and, then,  the Jost function. The latter $F_{\ell}(k)=f_{\ell}(k,\sigma)$
 for a potential which includes a hard core and $F_{\ell} (k)=\lim_{r \to 0} (-k r)^{\ell}
 f_{\ell}(k,r)/(2 \ell - 1) !! $ otherwise.
 We recall that $f_{\ell}(k,r) $ is defined by the Cauchy condition:
$$\lim_{r \to \infty} f_{\ell}(k,r)
\ \exp(-i k r) = i^{\ell} \ .$$
This gives us a Jost function which is analytic in the upper half plane
$\Im(k) > 0$.

For $\ell=0$,
let be $g(k,r)=f_0(k,r) \ \exp(-i k r)$. The function $g$ satisfies
$$
g(k,r) = 1 -\frac{i}{2 k} \ \int_r^{\infty} \left(e^{2 i k (r'-r)} -1
\right) \ V(r') \ g(k,r') \ dr' \ . \eqno (B1) $$

As usual\cite{Newt}, writing $g(k,r)$ as a series 
$g(k,r)=\sum g_n(k,r)$, we note that it is absolutely and uniformly
convergent when the potential $V$ satisfies
$\int_0^{\infty} r \ V(r) \ dr < \infty$.
We construct the $g_n$ by the recursive procedure
\beqa
g_0(k,r) & = & 1 \nonumber \\
g_n(k,r) & = & - \frac{i}{2 k} \ \int_r^{\infty} \left(e^{2 i k (r'-r)} -1
\right) \ V(r') \ g_{n-1}(k,r') \ dr'  \ n \geq 1 \ . \nonumber
\eeqa
We then obtain for a finite range potential,  infinitely  differentiable 
at the left of its cutoff denoted $a$, the terms
\beqa 
g_1(k,r) & = & \frac{i}{2 k} \int_r^a V(r') \ dr' \nonumber \\
 & & -\frac{1}{4 k^2} e^{2 i k
(a-r)} \left (V(a) -\frac{V'(a)}{2 i k} + \frac{V''(a)}{(2 i k)^2} + \cdots
\right) \nonumber \\
& & + \frac{1}{4 k^2} \left (V(r) -\frac{V'(r)}{2 i k} +
\frac{V''(r)}{(2 i k)^2} + \ \cdots \right) \ , \nonumber
\eeqa
and
\beqa
g_2(k,r) & = & - \frac{1}{8 k^2} \left( \int_r^a V(r') \ dr'\right)^2 
\nonumber \\
&& -\frac{1}{8  k^4} e^{2 i k
(a-r)} \left ( V(a)^2 -3  \frac{V'(a)\  V(a)}{(2 i k)} + \cdots
\right) \nonumber \\
& & -\frac{1}{8 i k^3} \left (V(r) \ \int_r^a V(r') \ dr'  +\frac{2 \ 
V^2(r)-V'(r) \
\int_r^a V(r') dr' }{2 i k} + \cdots \cdots  
\cdots \right) \nonumber \\
 & & - \frac{1}{16 k^4} e^{2 i k (a-r) } \left(V(a)-\frac{V'(a) 
 \ }{2 i k} + \cdots \right) \left(V(r)+\frac{V'(r) 
 \ }{2 i k} + \cdots \right) \nonumber \\
 && + \frac{i}{8 k^3} e^{2 i k (a-r)} 
 \left(V(a)-\frac{V'(a)}{2 i k} + \cdots \right) \int_r^a V(r') \ dr' 
 \nonumber \\
 & & + \frac{i}{8 k^3} \ \int_r^a V(r') \ \left(V(r')-\frac{V'(r')}{2 i k}
 +\cdots \right) dr'  \nonumber \\
 && +\frac{1}{16 k^4} \left(V(a)-\frac{V'(a) 
 \ }{2 i k} + \frac{V''(a)}{(2 i k)^2} \cdots \right) \left(V(a)+\frac{V'(a) 
 \ }{2 i k} + \frac{V''(a)}{(2 i k)^2} \cdots \right) \ , \nonumber
\eeqa
etc.. and
$$
g_3 (k,r) = \frac{1}{32  k^4} \ e^{2 i k (a-r)} \left[ \ V(a) 
\ \left(\int_r^a V(r') \ dr' \right)^2  + O(k^{-1}) \right] + O(k^{-3}) \ .
$$ 
To obtain the $g_n$'s, we have used partial integration on the factor
containing the exponential and differentiated the term containing the
potential.

If one approximates the function $g(k,r)$ by the sum
$s(k,r)=g_0+g_1+g_2 +g_3$,
the Jost function is approximated by $s(k,0)$, for a potential which is
finite at the origin, and by
$s(k,\sigma) \exp(i k \sigma) $ for a potential which includes
a hard core component.
We obtain, setting $M=\int_{\sigma}^a V(r') dr'$,
\beqa
e^{2 i k (a-\sigma) } & = & \frac{4 k^2}{V(a)} 
 \left[ 1 + \frac{i}{2 k}  \left(2 M-\frac{V'(a)}{V(a)} \right) \right.
 \nonumber \\
& +& \left. \frac{1}{4 k^2}\left(\frac{V''(a)}{V(a)}-\frac{V'(a)^2}{V(a)^2}
- 2 V(a) + 
2 M \frac{V'(a)}{V(a)} - 2 M^2  \right) \right] \ , \nonumber
\eeqa
as the condition  which will yield the poles of the ${\cal S}$ matrix,
 provided that $V(a) \ne 0$.

\newpage
\section*{Appendix C}
\noindent 

For higher waves we have to deal with the following  free Jost solution
$$
 w_{\ell}(k r) = i (-)^{\ell } \sqrt{ \frac{\pi}{2}  k r} \ H^{(1)}_{\ell +
 1/2}(k r) \ ,
\eqno (C1) $$ 
where $H^{(1)}_{\nu}$ is the Hankel function of  the first
kind of order $\nu$. We have
$$\lim_{r \to 0} (k r )^{\ell} \ w_{\ell} (k r) = 
 (-)^{\ell} \ (2 \ell-1) !! \ , $$
and
$$\lim_{r \to \infty} \exp(-i k r) \ w_{\ell}(k r) =i^{\ell} \ . $$
The $w_{\ell}$'s are simply given by \cite{Erde}
$$
w_{\ell} (kr) = i^{\ell} e^{i k r} P_{\ell}(kr) \ , \eqno (C2) $$
where the $P_{\ell}$'s denote the polynomial part of the Hankel, i.e. 
$$
P_{\ell} (k r) = \sum_{m=0}^{\ell} \frac{(l + m)!}{(l-m)! m!} \ \left(
\frac{i}{2 k r} \right)^m \ .
\eqno (C3) $$
We proceed in a manner similar to that used for $\ell=0$.
We introduce
$$
 g_{\ell}(k,r)= f_{\ell}(k,r) /w_{\ell}(kr) \ ,
\eqno (C4) $$
where $f_{\ell}(k,r)$ is the Jost solution having the appropriate behaviour
for $r$ tending to infinity
$$\lim_{r \to \infty} i^{\ell} \exp(-i k r) \ f_{\ell}(k,r) =1 \ .$$
Note that the $w_{\ell}$'s never vanish for $k$ real.
The function $g_{\ell}$ then satisfies
\beqa
g_{\ell} (k,r)& = & 1 - \frac{i}{2k} \ \int_r^{\infty} \  (e^{2 i k (r'-r) }
P_{\ell}(k r')^2 \ \frac{P_{\ell}(-k r)}{P_{\ell} (k r) } \nonumber \\
&- & P_{\ell} (k r' ) \ P_{\ell} (-k r') ) V(r') g_{\ell} (k,r' ) \ dr' \ .
\nonumber
\eeqa
We apply the previous procedure (see Appendix B)
which consists in using partial  integration for the factor containing the
exponential and differentiating the term containing the potential $V(r')$
multiplied by $P_{\ell}^2(kr')$.

When the potential includes an hard core component
the factors $(k r')^{-m} \ m >0$, occurring  in the polynomial $P_{\ell}$,
  are bounded by $(k \sigma)^{-m}$ and therefore goes to zero when $| k| $ 
  tends to infinity.

We then obtain  a formula similar to (15) but where   successive
derivatives of $P_{\ell}$ appear. This implies additional
$\ell$-dependent terms in the expansion in powers of $1/k$.

\newpage

For example, for the potential used before (hard core plus square well)
 we found,   
\beqa
e^{2 i k (a-\sigma) } & = & -\frac{4 k^2 (a-\sigma)^2}{A^2} 
\ \left[ 1 + \frac{i}{k (a-\sigma)} \left(\frac{x}{b (b+1)}
- A^2 \right) \right. \nonumber \\
& + & \left. \frac{1}{2 k^2 (a-\sigma)^2 } \left( 
-A^4 + 2 A^2 \frac{x}{b(b+1)} +
A^2 - x \frac{2 b^2 +x}{b^2 \ (b+1)^2} \right) \right. \nonumber \\
 & + & \left. \frac{i }{6 k^3 (a-\sigma)^3} \left( -\frac{3}{2} A^4 + A^6 - 
 3 A^4 \frac{x}{b (b+1)} + 3 x \frac{A^2}{b^2 (b+1)^2} (2 b^2 + x) \right.
 \right. \nonumber \\
 &+& \left.\left. \frac{x}{2 b^3 (1 + b)^3} 
 \ (-6 (1 + b)^3 + 18 b^3 + x (1 + 3 b -9 b^2) -2 x^2) \right) \right] \ , 
 \nonumber  
\eeqa
where $ A= 2 \pi (a-\sigma)/\Lambda^*; \ x=\ell (\ell +1) $ and $b=\sigma/
(a-\sigma)$.

When the potential has no hard core,
the factors $1/(k r)^m \ m>0$ occurring
in the $P_{\ell}$'s are no longer  bounded and we have to reason differently.

In fact, when $r$ tend to zero the term
$$\frac{P_{\ell}(-k r)}{P_{\ell} (k r) } \ , $$
tends  to $(-)^{\ell} $ and, in so  far as the leading term of $g_{\ell}$ is
concerned, we are left with
\beqa
g_{\ell} (k,r) & = & 1 - \frac{i}{2k} \ \int_r^{\infty} \  ( (-)^{\ell}
e^{2 i k (r'-r) } P_{\ell}(k r')^2 \frac{1 + i k r + \ldots }{1-i k r +
\ldots } \nonumber\\
& - & P_{\ell} (k r' ) \ P_{\ell} (-k r') ) V(r') \ dr' \ . \nonumber 
\eeqa
When $|k|$ is large, only the behaviour of the potential at its cutoff $a$
dominates
$$
g_{\ell} (k,r) = 1 - (-)^{\ell} \frac{V(a)}{4 k^2} e^{2 i k a} \ ,
\eqno (C5) $$
and the leading  asymptotic expression is given by solving
$$
e^{2 i k a} = (-)^{\ell} \frac{V(a)}{4 k^2} \ .
\eqno (C6) $$
We then recover the alternating sign, which depends on whether $\ell$ is even
or odd,  mentioned earlier by Nussenzveig\cite{Nuss}.
This dependence disappears when the potential incorporates a hard core.

\newpage
\section*{ Appendix D }
\vspace{.3cm}

We examine the low temperature expansion of the virial
for a pure hard core. We start from the equation (10), which for the
hard sphere reads:
$$
\left( B_2 \right)_{Boltz} = - 2^{1/2} \lambda_T^3 {\cal N} \
 \sum_{\ell} (2 \ell +1)
\left[ -\frac{\sigma}{ \sqrt{ 2 } \lambda_T}
+ \frac{1}{2}  \sum_n \exp(- \frac{\lambda_T^2}{2 \pi}
k_{n,\ell}^2) 
\erfc(i \frac{ \lambda_T}{\sqrt{2 \pi}}  k_{n,\ell}  )   \ \right] \ .
\eqno(D1)
$$
The equation (D1) involves the function $\exp(-z^2) \ 
\erfc(i z  ) $, where $ z=\lambda_T \  k_{n,\ell}/\sqrt{2 \pi}$. This latter 
 has the asymptotic expression for $\lambda_T/\sigma$ (or equivalently $z$ ) 
 large
$$
\exp(-z^2) \ \erfc(i z) = -\frac{i}{\sqrt{\pi} \ z} \left[ 1 +
\sum_{j=1}^{\infty} \frac{(2 j-1)!!}{(2 z^2)^j} \right] \ .
\eqno(D2)
$$
Incorporating (D2), written  for $z=\lambda_T \  k_{n,\ell}/\sqrt{2 \pi} $,
into (D1) we have:
\begin{eqnarray}
\left( B_2 \right)_{Boltz}& = & - 2^{1/2} \lambda_T^3 {\cal N} \
\left[-\frac{\sigma}{ \sqrt{ 2 } \lambda_T}
 +   \sum_{\ell \ne 0} (2 \ell +1)
\left( -\frac{\sigma}{ \sqrt{ 2 } \lambda_T} \right.\right. \nonumber\\
&-&\left.\left. \frac{i}{ \sqrt{ 2 } \lambda_T} 
\ \sum_{n=1}^{\ell} \left[ \frac{1}{k_{n,\ell} }  + \sum_{j=1}^{\infty}
 \frac{(2 j-1)!!}{k_{n,\ell}^{2 j +1}} \ \frac{\pi^j}{\lambda_T^{2 j}}
 \right]  \right) \ \right] \ . \nonumber
\end{eqnarray}
In the previous equation  use is made  of the property  for  hard spheres
the ${\cal S}$
matrix has no poles for $\ell=0$ and  exactly $\ell$ poles 
for $\ell \ne 0$.

The expression for the virial, divided by its classical limit,
as in (30) reads:
\begin{eqnarray}
\left( B_2^* \right)_{Boltz}& = & \frac{3}{2 \pi} 
\left(\frac{\lambda_T}{\sigma}\right)^2
\left[1 +   \sum_{\ell \ne 0} (2 \ell +1) \left(1 + 
\sum_{n=1}^{\ell} \frac{i}{(k_{n,\ell} \ \sigma) } \right.\right. \nonumber\\
\ & + & \left.\left. \sum_{j=1}^{\infty}
 (2 j-1)!! \ (-)^j \ \pi^j  \left(\frac{\sigma}{\lambda_T} \right)^{2 j}
\ \sum_{n=1}^{\ell} \left(\frac{i}{(k_{n,\ell} \  \sigma)}\right)^{2 j +1}
 \   \right) \ \right] \ . \nonumber
\end{eqnarray}

The calculation of the virial requires the knowledge of the sums
$$
S_{j,\ell}=\sum_{n=1}^{\ell} \left(\frac{i}{(k_{n,\ell} \ \sigma)} \right)^{2
j + 1} \qquad\quad j \leq 1 \ .
\eqno(D3)
$$

The poles $k_{n,\ell}$ of the ${\cal S}$ matrix are the zeros
of the polynomial part of the Hankel function (Eq.(C3) for $r=\sigma$). 
Introducing $x_{n,\ell}= i/( k_{n,\ell} \ \sigma)$, these latter are 
roots  of the polynomial
$$
P(x)=\sum_{m=0}^{\ell} a_{m,\ell}  \ x^m \ ,  
\eqno (D4) $$
with
$$
a_{m,\ell}=\frac{(l + m)!}{2^m \ (l-m)! m!} \ .
\eqno (D5) $$

The sums $S_{j,\ell}$, Eq.(D3), are given by
$$
S_{j,\ell}=\sum_{n=1}^{\ell} x_{n,\ell}^{2 j + 1} \qquad\quad j \leq 1 \ ,
\eqno(D6)
$$
in terms of the roots of the polynomial Eqs. (D4,D5).

They obey the recursion formula
$$
 S_{1,\ell}  = - \frac{a_{\ell-1,\ell}}{a_{\ell,\ell}} 
      $$
$$S_{j,\ell}= -\sum_{m=1}^{j-1} \frac{a_{\ell-m,\ell}}{a_{\ell,\ell}} \
S_{m,\ell} - j \ \frac{a_{\ell-j,\ell}}{a_{\ell,\ell}} \qquad\quad j \geq 2 \ .
\eqno(D7)
$$
\indent From (D5) and (D7) we find,  after calculation, 

\begin{center}
\begin{tabular}{|r|r|r|r|r|r|r|}
\hline
 $\ell$ & $S_{1,\ell}$ & $S_{3,\ell}$
 & $S_{5,\ell}$ & $S_{7,\ell}$ & $ S_{9,\ell}$
 & $S_{11,\ell}$ \\
\hline
 1 &  -1  &  -1  &  -1      &  -1 &  -1  &  -1  \\
 2 &  -1 & 0  & $1/9$ & $1/27 $  &  0  & $-1/243$    \\
 3 &  -1 &  0    & 0 & $-1/225  $  &  $-1/1125$ &
  $-1/16875$ \\
 4 &  -1 &  0    &  0  & 0 &  $1/11025 $   &
 $ 1/77175$ \\
 5 &  -1 &  0    &  0      &     0  & 0 & $-1/893025 $   \\
\hline
\end{tabular}
\end{center}

The sums $S_{2 j+1,\ell} , j \ne 0 $ are zero for $\ell \geq j+1 $.

Rewriting $\left( B_2^* \right)_{Boltz}$ in terms of the $S_{j,\ell}$'s
$$
\left( B_2^* \right)_{Boltz} = \frac{3}{2 \pi} \ \frac{\lambda_T^2}{
\ \sigma^2}
\left[ 
1 +   \sum_{j=1}^{\infty} (2 j-1)!! \  (-)^j \ 
\pi^j \left(\frac{\sigma}{\lambda_T} \right)^{2 j}
\ \sum_{\ell \ne 0} (2 \ell+1) \ S_{2 j +1,\ell}
 \ \right] \ ,
\eqno(D8)
$$
and, taking into account the results depicted in the table, we have
\beqa
\left( B_2^* \right)_{Boltz} & = & \frac{3}{2 \pi } \
 \frac{\lambda_T^2}{ \sigma^2}
\left[ 
1 +  3 \pi   \left(\frac{\sigma}{\lambda_T} \right)^{2 }
- \frac{22}{3} \ \pi^2 \ \left(\frac{\sigma}{\lambda_T} \right)^{4 }
+ \frac{1921}{45} \pi^3 \left(\frac{\sigma}{\lambda_T} \right)^{6 }
-\frac{165673}{525} \pi^4 \left(\frac{\sigma}{\lambda_T} \right)^{8 }
\right. \nonumber\\
&+& \left. \frac{472102277}{165375} \ \pi^5 
\left(\frac{\sigma}{\lambda_T} \right)^{10 } +\ldots 
 \ \right] \ \ . \nonumber 
\eeqa
We thus recover the first terms extracted by BLK.

\end{document}